\documentclass[aps,pre,reprint,superscriptaddress,amsmath,amssymb,floatfix]{revtex4-1}
\usepackage{graphicx}
\usepackage{bm}
\usepackage{amsmath}  

\usepackage{color,ulem,soul}

\soulregister\cite7
\soulregister\ref7

\begin{document}
	
	
\title{Emergence of wrinkles during the curing of coatings}
	
\author{Michiko Shimokawa}
\author{Hikaru Yoshida}
\author{Takumi Komatsu}
\affiliation{Fukuoka institute of Technology, Wajiro-higashi, Higashi-ku,
		Fukuoka 811-0295, Japan}
	
\author{Kazue Kudo}
\affiliation{Department of Computer Science, Ochanomizu University,
	2-1-1 Ohtsuka, Bunkyo-ku, Tokyo 112-8610, Japan}

\date{\today}
	
\begin{abstract}
Wrinkles often emerge on a paint layer when a second coat of paint is applied on an already-coated substrate. 
Wrinkle formation occurs when the first layer absorbs organic solvent from the second layer. 
We set up experiments to mimic the double-coating process, focusing on the interaction between a paint layer and an organic solvent. 
In the experiments, we investigated the characteristic wavelengths of the wrinkles and the time of wrinkle emergence. 
We propose a simple model to explain (1) the buckling induced by swelling of the layer due to absorption of the organic solvent and
(2) delamination of the layer from the substrate.
 A linear stability analysis of the model provides a relationship between the wavelengths and the characteristic timescale 
 that agrees reasonably well with the data obtained from our experiments. 
 Our results indicate that compression of the layer due to swelling and delamination are both important factors in the formation of wrinkles.
\end{abstract}
	
\maketitle
\section{Introduction}
Double coating, in which paint is applied over an already-coated substrate, is often used to avoid unevenness in the paint layers. 
In spite of double coating, however, wrinkles sometimes emerge in the drying process, 
if (1) the elapsed time between the first and second coatings is too short, or (2) the coating thickness is too large~\cite{Basu_POC2005,Basu_JAPS}. 
The formation of wrinkles has been studied in many fields, such as engineering, 
material science, chemistry, and physics~\cite{Thin_film}. 
The fundamental process of wrinkle formation, however, is not yet fully understood.
Below, we focus on case (1) above and investigate the formation of wrinkles from the viewpoint of the mechanical stability of the paint layer.

In a double-coating process, two layers of paint are produced by the first and second coatings. 
Deformation of the first layer, underlying the second layer, leads to the formation of wrinkles observed at the surface of the second layer. 
A resin paint, which includes a polymer and an organic solvent, is often used in the painting process. 
The mechanism of wrinkle formation by a resin paint is thought to be as follows: 
A polymerization reaction proceeds in the layer after the first application, and the strength of the layer increases as it cures~\cite{handbook}. 
When a second coating is applied, the organic solvent, which is an ingredient of the second coating of paint, penetrates into the first layer. 
Exposure to the organic solvent causes the polymerized first layer to swell~\cite{Burrell}. 
The first layer is easily swollen when the elapsed time between the first and second coatings is too short, because of incomplete polymerization of the first layer~\cite{Gel}. 
The swelling induced by absorption of the solvent thus causes deformation of the first layer, 
and wrinkle formation at the surface of the second layer is due to the resulting deformation of the underlying first layer. 
Most previous experiments on double coating have focused on the top (second) layer rather than the first layer~\cite{Basu_POC2005, Basu_shiwa}.

In this paper, we describe an experiment in which an organic solvent is applied to the surface of the first layer. 
We can then observe the deformation of the first layer directly in the experiment, without the complicating effects of the deformation and polymerization of the second layer. 
The emergence of wrinkles in this experiment is due solely to the deformations caused by the instability of the first layer. 
We investigate the characteristic length scales, i.e., the wavelengths, of the wrinkles, together with the characteristic timescale that characterizes wrinkle formation. 
The characteristic lengths and the timescale depend on the elapsed time $T$ between the application of the first coating and the application of the organic solvent. 
We propose a simple model that includes both the effects of buckling due to the swelling of the layer and delamination of the layer from the substrate. 
The model yields a relationship between the wavelengths and the characteristic timescale, which is similar to that obtained from the experiments. 
The results show that the swelling of the layer and its delamination from the substrate cause the instability of the layer that leads to the emergence of wrinkles. 
In the following sections, we discuss the experimental data in some detail and consider the process of wrinkle formation by means of the model.

\section{Experimental method}
\begin{figure}[htb]
 \begin{center}
	\includegraphics[width=7cm,clip]{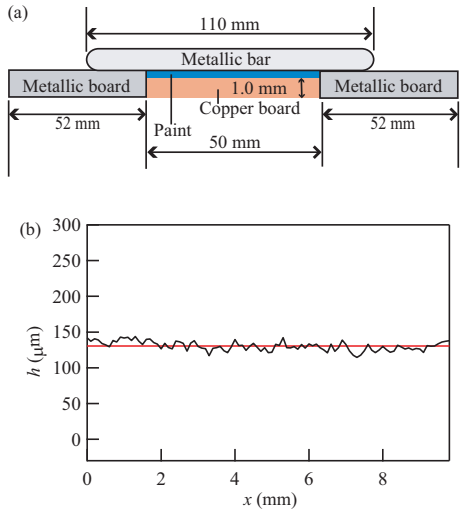}
 \end{center}
 \caption{\label{fig:1} (Color online)
(a) Schematic drawing of the experimental setup for the application of a paint layer. 
Only the coated copper board is kept in a constant-temperature oven for several hours after the first coating. 
(b) The surface height $h$ measured after painting.
 }
\end{figure}

We used a copper board of 5.0 cm $\times$ 5.0 cm square and 1.0 mm thick as the substrate for painting. 
To control the thickness of the paint layer, we placed two metallic boards facing each other on opposite sides of the copper board, as shown in Fig.~\ref{fig:1}(a). 
The metallic boards are of equal thickness and are slightly thicker than the copper board. 
We applied a phthalic resin paint (Rubicon1000, No. 837, ISHIKAWA PAINT) to the copper board using a syringe (SS-20ESZ, TERMO), 
and we spread the paint across the copper board using a metallic bar, producing a layer of relatively uniform thickness. 
We measured the thickness $h$ of the paint layer using a laser displacement meter (LT9010M, KEYENCE). 
As shown in Fig.~\ref{fig:1}(b), we found that the layer has a nearly uniform thickness with an average value $h = 130 \pm 6$ $\mu$m. 
The coated board was then placed in a constant-temperature oven (NEXAS OFX-70, ASONE) at  $30^\circ$C for a time $T$. 
After the time $T$, we applied a $0.02$ cm$^3$ drop of xylene, which is the organic solvent in the paint, on the coated layer. 
The surface of the layer was photographed with a digital camera (Canon EOS Kiss X4, EF-S 18-55IS) 10 min after the xylene application. 
It is easy to observe the deformations of the paint layer, since xylene is clear and colorless.

\section{Experimental Results}
\subsection{Buckles formation}
\label{sec:3A}

\begin{figure}[htb]
 \begin{center}
  \includegraphics[width=7cm,clip]{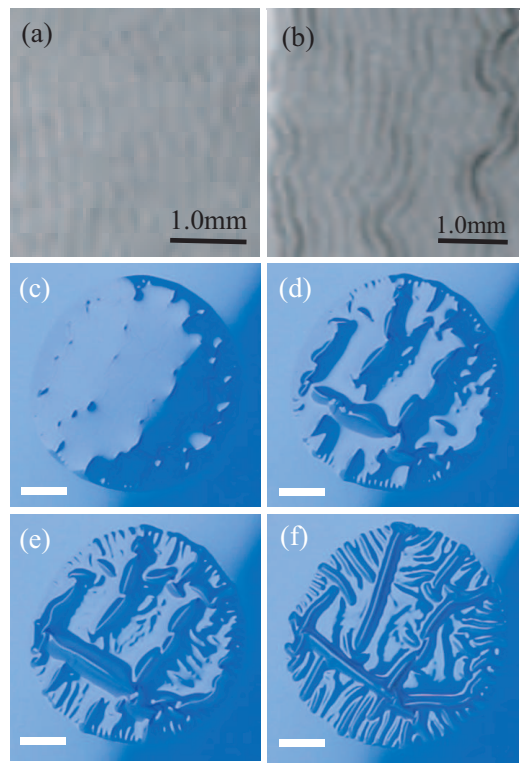}
 \end{center}
 \caption{\label{fig:2-2} (Color online)
Deformations of the paint layer at (a) $t = 58$ s, (b) 63 s, (c) 90 s, (d) 120 s, (e) 150 s, and (f) 300 s, 
where $t = 0$ is the time when the organic solvent is applied to the layer. 
These photographs all apply to experiments for which the curing time $T = 24$ h. 
The solid lines in panels (a) and (b) are 1.0-mm-scale bars, and those in panels (c)--(f) are 3.0-mm-scale bars. 	
}
\end{figure}

Figure~\ref{fig:2-2} shows the deformations of the paint layer at the following times $t$ after application of the organic solvent: 
(a) $t =$ 58 s, (b) 63 s, (c) 90 s, (d) 120 s, (e) 150 s, and (f) 300 s, all for experiments at the fixed time $T = 24$ h after the application of the paint layer. 
The drop of organic solvent spreads into a circular shape approximately 11 mm in diameter. 
Short-scale wrinkles first emerge at $t = 58$ s [Fig.~\ref{fig:2-2}(a)]. 
Shortly after that, at $t = 63$ s, larger-scale wrinkles appear [Fig.~\ref{fig:2-2}(b)], 
and small bumps appear randomly at $t = 90$ s [Fig.~\ref{fig:2-2}(c)]. 
The bump amplitudes are much larger than those of wrinkles observed at earlier times. 
The amplitudes of the bumps increase with $t$, and delamination of the layer from the copper board occurs. 
In the process, some bumps coalesce with other bumps [Figs.~\ref{fig:2-2}(d), (e)]. 
The coalescence repeats, and buckles emerge, as shown in Fig.~\ref{fig:2-2}(f). 
The pattern of buckles does not change after $t = 300$ s.
	
\begin{figure}[htb]
 \begin{center}
  \includegraphics[width=7cm,clip]{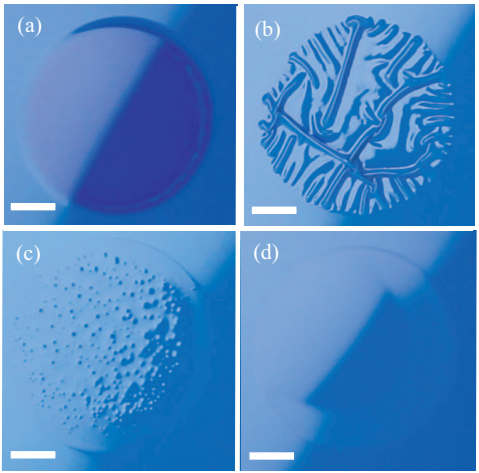}
 \end{center}
 \caption{\label{fig:3} (Color online)
 Deformations of layers obtained in experiments for (a) $T = 1$ h, (b) 24 h, (c) 56 h, and (d) 64 h, 
 where $T$ is the time elapsed between the application of the paint layer and the application of the drop of organic solvent. 
 These images were taken at $t = 600$ s after the application of the organic solvent. 
 Solid lines in the photos are 3.0-mm-scale bars.	
 }
\end{figure}

We focus here on the deformations of the paint layer that occur in experiments for several different values of the curing time $T$.  
Figure~\ref{fig:3} shows snapshots for (a) $T = 1$ h, (b) 24 h, (c) 56 h, and (d) 64 h. 
These images were all obtained at $t =10$ min. 
Buckles emerge only in experiments for $T = 24$ h [Fig.~\ref{fig:3}(b)]. 
The paint layer is melted by the organic solvent in experiments for $T = 1$ h [Fig.~\ref{fig:3}(a)]. 
Several bumps appear in experiments for $T = 56$ h [Fig.~\ref{fig:3}(c)], 
but they vanish at $t = 30$ min, and a layer with a smooth surface remains. 
In experiments for $T = 64$ h, the surface of the layer remains smooth and does not change with time [Fig.~\ref{fig:3}(d)]. 
The experiments for $T = 56$ and 64 h both result in smooth surfaces, even after the application of the organic solvent, 
but the processes by which the smooth surfaces are produced are different. 
These results show that buckles emerge only for a limited range of the curing times $T$. 
This behavior is similar to the results obtained in previous experiments with double coatings \cite{Basu_POC2005,Basu_shiwa}.

\subsection{Characteristic spatial scales and timescales for the formation of wrinkles}

\begin{figure*}[htb]
 \begin{center}
  \includegraphics[width=14cm,clip]{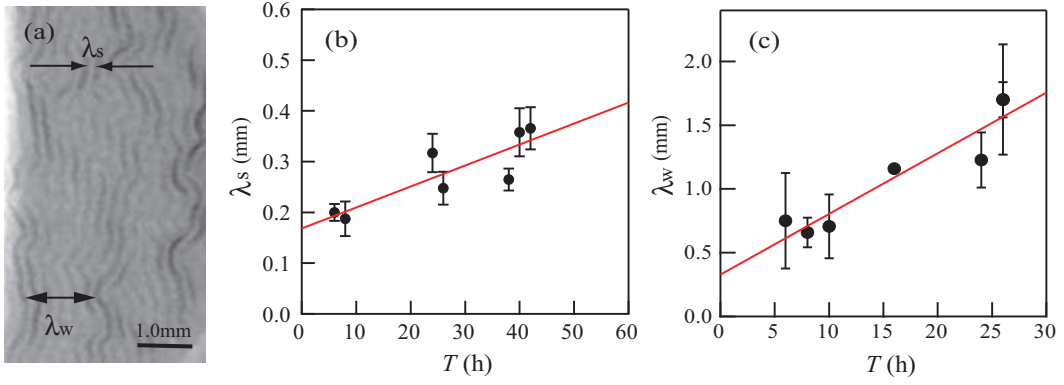}
 \end{center}
 \caption{\label{fig:7-3} (Color online)
(a) Snapshot taken at $t =$ 63 s in the experiment with elapsed time $T = 24$ h. 
The short-scale wavelength  $\lambda_{\rm s}$ of the small wrinkles and the wavelength  $\lambda_{\rm w}$ of the larger-scale wrinkles are indicated. 
The solid line in the photo shows a 1.0-mm-scale bar. 
Panels (b) and (c) show the quantities  $\lambda_{\rm s}$ and  $\lambda_{\rm w}$ obtained from our experiments for several different values of $T$. 
The closed circles are the experimental data, and the solid lines in (b) and (c) are the fitted lines given by Eqs.~\eqref{eq:lambda_s} and \eqref{eq:lambda_w}, respectively. 	
 }
\end{figure*}

Figure~\ref{fig:7-3}(a) shows the short-scale wavelength $\lambda_{\rm s}$ and the wavelength $\lambda_{\rm w}$ of larger-scale wrinkles 
obtained from experiments with $T = 24$ h, which were pointed out in Sec.~\ref{sec:3A}. 
Small structures with wavelengths $\lambda_{\rm s}$ appear first, and wrinkles with wavelengths $\lambda_{\rm w}$ appear subsequently. 
The quantity $\lambda_{\rm w}$ is the maximum wavelength observed before the wrinkles coalesce. 
Figures~\ref{fig:7-3}(b) and (c) show the values of $\lambda_{\rm s}$ and  $\lambda_{\rm w}$, respectively, 
in experiments for several different values of $T$. 
They can be fitted by the linear functions

\begin{align}
\lambda_{\rm s} &= c_{\rm s1}T + c_{\rm s2},
\label{eq:lambda_s}
\\
\lambda_{\rm w} &= c_{\rm w1} T + c_{\rm w2} 
\label{eq:lambda_w} 
\end{align}
with the fitting parameters  $c_{\rm s1}=0.41\times10^{-2}$,  $c_{\rm s2}=0.17$, $c_{\rm w1}=0.44\times10^{-1}$, and $c_{\rm w2}=0.37$. 
Using Eqs.~\eqref{eq:lambda_s} and \eqref{eq:lambda_w}, we can determine $\lambda_{\rm s}$ and $\lambda_{\rm w}$ for any value of $T$.

\begin{figure}[htb]
 \begin{center}
  \includegraphics[width=7cm,clip]{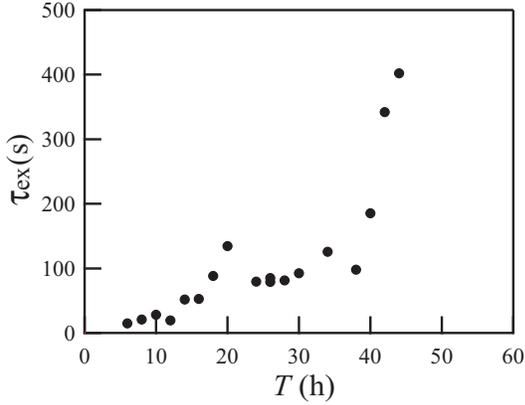}
 \end{center}
 \caption{\label{fig:tau} (Color online)
Relationship between $T$ and $\tau_{\rm ex}$, 
where $\tau_{\rm ex}$ is the time elapsed between the application of the organic solvent and the appearance of bumps of 0.2 mm diameter. 
The closed circles are the experimental data.
 }
\end{figure}

Next, we investigate the characteristic timescale $\tau_{\rm ex}$, 
which turns out to be inversely related to the growth rate of the wrinkles. 
We define $\tau_{\rm ex}$ as the time elapsed between the application of the organic solvent and the appearance of bumps of 0.2 mm diameter. 
As shown in Fig.~\ref{fig:tau}, $\tau_{\rm ex}$ increases with $T$. 
The timescale $\tau_{\rm ex}$ is larger than the times at which $\lambda_{\rm s}$ and $\lambda_{\rm w}$ are measured. 
After the initial growth of patterns with wavelengths $\lambda_{\rm s}$ and $\lambda_{\rm w}$, 
the coalescence of wrinkles is caused by nonlinear effects. 
Coalescence leads to a change in the characteristic length of pattern deformation. 
The time when such a change occurs is proportional to the timescale $\tau_{\rm ex}$~\cite{tau}.

\section{Theoretical model}

We next introduce a simple model for the buckling that is
observed in our experiments.
Consider the coating of paint to be an elastic thin film that is
attached adhesively to a solid substrate.
Because of the absorption of the organic solvent, the elastic film swells, producing compression stress.  
Suppose that the elastic film exists on a flat
substrate, whose surface corresponds to the $x_1$-$x_2$ plane, and  
let the $x_3$ axis be normal to the surface of the substrate.
The total energy $F_{\rm tot}$ of this system consists of the elastic
strain energy in the film and the interfacial traction energy between
film and substrate:
\begin{align}
 F_{\rm tot}=\iint (f_{\rm film}+f_{\rm int})
 \; d x_1 d x_2,
 \label{eq:F_tot}
\end{align}
where $f_{\rm film}$ and $f_{\rm int}$ are the energies per unit area of
the film and the interface, respectively.

According to the F\"{o}ppl--von K\'{a}rm\'{a}n plate theory, the 
elastic energy per unit area in a film of thickness $h$ is given 
by \cite{tau,Faou,Pan1,Ni,Pan2}
\begin{align}
 f_{\rm film} &= \int_{-h/2}^{h/2}\frac12 \sigma^{\rm el}_{\alpha\beta}
 \varepsilon^{\rm el}_{\alpha\beta} \; dx_3,
 \label{eq:film} \\
 \sigma^{\rm el}_{\alpha\beta} &= \frac{2\mu}{1-\nu}\left[
 (1-\nu)\varepsilon^{\rm el}_{\alpha\beta}
 +\nu\varepsilon^{\rm el}_{\gamma\gamma}\delta_{\alpha\beta}\right],
 \label{eq:sg_el}\\
 \varepsilon^{\rm el}_{\alpha\beta} &= e_{\alpha\beta}
 -x_3\frac{\partial^2 w}{\partial x_\alpha\partial x_\beta}.
 \label{eq:ep_el}
\end{align}
Greek subscripts refer to in-plane coordinates $x_1$ or $x_2$, 
and repeated Greek subscripts indicate summation over indices $1$ and $2$.  
The parameters $\mu$ and $\nu$ are the shear modulus and 
Poisson ratio of the film, respectively. 
Mid-plane displacements in the in-plane and $x_3$ directions are
denoted by $u_\alpha$ and $w$, respectively.
Supposing the film to be under equibiaxial stress, 
we take the initial in-plane strain to be 
$\varepsilon_0\delta_{\alpha\beta}$. Then, 
\begin{align}
e_{\alpha\beta} = \frac12\left(
\frac{\partial u_\alpha}{\partial x_\beta}
+ \frac{\partial u_\beta}{\partial x_\alpha} \right)
+ \frac12 \frac{\partial w}{\partial x_\alpha}
\frac{\partial w}{\partial x_\beta}
	- \varepsilon_0\delta_{\alpha\beta}.
	\label{eq:e.1}
\end{align}

To express the interfacial traction energy between the film and 
substrate, we use the cohesive zone model~\cite{c-zone1,c-zone2,c-zone3}.
The interfacial energy per unit area is then
\begin{align}
	f_{\rm int} = \int_0^\zeta T_{\rm n}(z)\; dz,
	\label{eq:int}
\end{align}
where $\zeta$ is the distance between the substrate and film,
and $T_{\rm n}$ is the normal traction. 
When the film thickness is constant, $\zeta=w$.
We represent the normal traction as
\begin{align}
T_{\rm n}(\zeta) = \Gamma_{\rm n}\zeta
\exp\left(-\frac{\zeta}{\delta_{\rm n}}\right), 
	\label{eq:Tn}
\end{align}
where $\Gamma_{\rm n}\equiv\gamma_{\rm n}/\delta_{\rm n}^2$.
The parameters $\gamma_{\rm n}$ and $\delta_{\rm n}$ are the normal 
interfacial toughness and the
characteristic length of a normal displacement jump, respectively.

The total energy $F_{\rm tot}$ is thus expressed in terms of the displacements $u_\alpha$
and $w$. Equilibrium states must satisfy
$\delta F_{\rm tot}/\delta w=0$ and $\delta F_{\rm tot}/\delta u_\alpha=0$. 
However, instead of solving $\delta F_{\rm tot}/\delta w=0$, we employ the time-dependent
Ginzburg--Landau equation, which is often used in dynamical systems,
\begin{align}
	\frac{\partial w}{\partial t} = -\eta\frac{\delta F_{\rm tot}}{\delta w},
	\label{eq:GL}
\end{align}
where $\eta$ is a constant related to the characteristic relaxation time.
Scaling all lengths by $h$, times by $h/(\mu\eta)$,
the nondimensional equation
and $\gamma_{\rm n}$ by $h\mu$ in Eq.~\eqref{eq:GL}, we obtain
\begin{align}
	\frac{\partial w}{\partial t} = -\frac{1}{6(1-\nu)}\nabla^2\nabla^2w
	+ \frac{\partial N_\beta}{\partial x_\beta} - T_{\rm n},
	\label{eq:dwdt}
\end{align}
where the variables are dimensionless, $T_{\rm n}$ is the nondimensional
form of Eq.~\eqref{eq:Tn}, and
\begin{align}
	N_\beta = \frac{2}{1-\nu}\left[
	(1-\nu)e_{\alpha\beta} + \nu e_{\gamma\gamma}\delta_{\alpha\beta}
	\right] \frac{\partial w}{\partial x_\alpha}.
\label{eq:N}
\end{align}
The in-plane displacements $u_\alpha$ included in $e_{\alpha\beta}$ are
obtained from the equation $\delta F_{\rm tot}/\delta u_\alpha=0$.

\section{Linear stability analysis}
A linear stability analysis of Eq.~\eqref{eq:dwdt} provides some insight
into the condition of buckling.
Linearizing Eq.~\eqref{eq:dwdt} around $w=0$, and taking the Fourier transform
of the linearized equation, we obtain
\begin{align}
 \frac{\partial \tilde{w}(k)}{\partial t} = g(k)\tilde{w}(k),
 \label{eq:FT}
\end{align}
where $\tilde{w}$ is the Fourier transform of $w$, and $k$ is the wavenumber.
The linear growth rate $g$ is given by
\begin{align}
 g(k)=-\frac{1}{6(1-\nu)}[k^2 - 6(1+\nu)\varepsilon_0]^2
+ \frac{6(1+\nu)^2}{1-\nu}\varepsilon_0^2
 -\Gamma_{\rm n}.
\label{eq:g}
\end{align} 
Unstable modes, which cause deformations of the layer, appear when
$g(k)>0$; in other words,
\begin{equation}
 \Gamma_{\rm n} <
  \frac{6(1+\nu)^2}{1-\nu}\varepsilon_0^2.
  \label{eq:ep0}
\end{equation}
This equation shows that wrinkles emerge above a certain threshold of stress. 
The existence of the threshold is consistent with
the experimental results shown in Fig.~\ref{fig:3}, which indicate that buckles emerge under an upper limit of $T$, 
since $\Gamma_{\rm n}$ and $\varepsilon_0$ depend on $T$.
Equation~\eqref{eq:g} shows that the wavenumber
of the fastest-growing mode is
\begin{align}
k_{\rm f}= \sqrt{6(1+\nu)\varepsilon_0}.
\label{eq:kf}
\end{align}
The growth rate of the fastest-growing mode is inversely proportional to the timescale
\begin{align}
 \tau_{\rm f} = \left[\frac{6(1+\nu)^2}{1-\nu}\varepsilon_0^2 
-\Gamma_{\rm n}\right]^{-1}.
\label{eq:tau_peak}
\end{align}
Suppose that the growth rate of a certain unstable mode $k$ is $g(k)=C$ $>0$.
When $k<k_{\rm f}$, we have
\begin{align}
k^{2} &= 6(1+\nu)\varepsilon_0-\sqrt{6(1-\nu)(\tau_{\rm f}^{-1}-C)}
\nonumber \\
&= {k_{\rm f}}^2-\sqrt{6(1-\nu)(\tau_{\rm f}^{-1}-C)},
\label{eq:k}
\end{align} 
where $C$ is a constant.
Equation~\eqref{eq:k} leads
\begin{align}
\tau_{\rm f}=\frac{6(1-\nu)}{(k_{\rm f}^2-k^2)^2+C'}
\propto\left[ 
\left( \frac{1}{\lambda_{\rm f}^2} - \frac{1}{\lambda^2}\right )^{2}
+C'\right]^{-1},
\label{eq:tau1_1}
\end{align} 
where $k_{\rm f}=2\pi/\lambda_{\rm f}$, $k=2\pi/\lambda$, and $C'$ 
is a constant. 
We also take $\nu$ as a constant, as done in previous work~\cite{tau,Faou,Pan1,Ni,Pan2}.

By using Eq.~\eqref{eq:tau1_1}, we here examine the validity of our model.
We assume
that $\lambda_{\rm f}$ and $\lambda$ in Eq.~\eqref{eq:tau1_1} correspond to 
$\lambda_{\rm s}$ and $\lambda_{\rm w}$ in Fig.~\ref{fig:7-3}, respectively.
This assumption implies that structures with wavelengths $\lambda_{\rm s}$ and $\lambda_{\rm w}$ 
appear in the linear-instability region and that $\lambda_{\rm s}$ and $\lambda_{\rm w}$ 
correspond to unstable modes of pattern formation.
We also assume that $\tau_{\rm ex} \propto \tau_{\rm f}$~\cite{tau}.

\begin{figure}[htb]
 \begin{center}
  \includegraphics[width=7cm,clip]{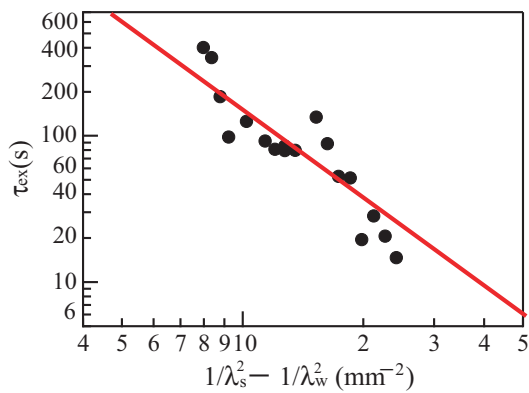}
 \end{center}
 \caption{\label{fig:fitting} (Color online)
 The relationship between  $\tau_{\rm ex}$ and $(1/\lambda_{\rm s}^2)-(1/\lambda_{\rm w}^2)$, 
 where $\tau_{\rm ex}$ is the time between the application of the organic solvent 
 and the appearance of bumps of 0.2 mm diameter.
The quantities $\lambda_{\rm s}$ and $\lambda_{\rm w}$ are 
the wavelengths of the small wrinkles that first appear 
and the maximum wavelength observed before the wrinkles coalesce, respectively.
 The closed circles are the experimental results, and the solid line is the fit from Eq.~\eqref{eq:tau1_1}. 
 }
\end{figure}

The closed circles in Fig.~\ref{fig:fitting} show the relationship between $\tau_{\rm ex}$ and 
$(1/\lambda_{\rm s}^2)-(1/\lambda_{\rm w}^2)$ obtained
in our experiments for several values of $T$.
Values of $\lambda_{\rm s}$ and $\lambda_{\rm w}$ are obtained using Eqs.~\eqref{eq:lambda_s} and ~\eqref{eq:lambda_w}.
The experimental data agree reasonably closely with the line given by Eq.~\eqref{eq:tau1_1}, which is shown as a solid line.
The value of $C'$ in Eq.~\eqref{eq:tau1_1} nearly vanishes, and
the fitted line is a power law with exponent $-2$.

The agreement between the experimental results and the fitted line in Fig.~\ref{fig:fitting} illustrates the validity of our model in the linear-growth regime.
The linear stability analysis thus shows that the emergence of wrinkles depends on both the initial strain caused by volume expansion and the normal traction.
Thus, the experimental results imply
that (1) buckling due to the volume expansion of the paint layer and (2) delamination of the layer from the substrate are both important factors for the emergence of wrinkles. 

\section{Numerical simulation}

Numerical simulation is useful for demonstrating that our simple model does reproduce buckling of the film.
We employ a spectral method for simulations.
The Fourier transform of Eq.~\eqref{eq:dwdt} is
\begin{align}
 \frac{\partial\tilde{w}}{\partial t}=-Dk^4\tilde{w}-ik_\beta\tilde{N}_\beta
-\tilde{T}_{\rm n},
\label{eq:dwdt.1}
\end{align}
where $D=1/[6(1-\nu)]$ and $\tilde{N}_\beta$ and $\tilde{T}_{\rm n}$ are the Fourier transforms of Eq.~\eqref{eq:N} and of the normal traction $T_{\rm n}$, respectively. 

The nonlinear term $N_\beta$ includes derivatives of $u_\alpha$. 
By using the condition $\delta F_{\rm tot}/\delta u_\alpha=0$, the Fourier transform of $u_\alpha$ can be written as
\begin{align}
 \tilde{u}_\alpha = \tilde{G}_{\alpha\beta}\tilde{\rho}_\beta,
\label{eq:u_a}
\end{align}
where
\begin{align}
 \tilde{G}_{\alpha\beta} &= \frac{1}{1-\nu}\left(
\frac{\delta_{\alpha\beta}}{k^2} 
- \frac{1+\nu}{2}\frac{k_\alpha k_\beta}{k^4}\right),
\label{eq:G_ab}
\\
\tilde{\rho}_\alpha &= \iint \left[
(1+\nu)\frac{\partial w}{\partial x_\gamma}
\frac{\partial^2 w}{\partial x_\alpha \partial x_\gamma}
+ (1-\nu)\frac{\partial w}{\partial x_\alpha}\nabla^2 w
\right]
\nonumber\\
&\quad\quad 
e^{i\bm{k}\cdot\bm{r}}dx_1dx_2.
\label{eq:rho_a}
\end{align}
Using Eqs.~\eqref{eq:u_a}, \eqref{eq:G_ab}, and \eqref{eq:rho_a}, we can rewrite Eq.~\eqref{eq:e.1} in the following form~\cite{Ni,Pan2},
\begin{align}
  e_{\alpha\beta} = \frac12
  &\int_{k\neq 0}\left[
-i(k_\beta\tilde{G}_{\alpha\gamma} + k_\alpha\tilde{G}_{\beta\gamma})
\tilde{\rho}_\gamma
\right]\frac{e^{-i\bm{k}\cdot\bm{r}}}{(2\pi)^2}d^2k
\nonumber\\
&+ \frac12\frac{\partial w}{\partial x_{\alpha}}
\frac{\partial w}{\partial x_{\beta}}
- \varepsilon_0\delta_{\alpha\beta}.
\label{eq:e.2}
\end{align}

In numerical simulations, we use the modified normal traction,
\begin{align}
 T_{\rm n}(\zeta)=
\begin{cases}
\Gamma_{\rm n}\zeta\exp(-\zeta/\delta_{\rm n}) & \zeta\ge 0, \\
\Gamma_{\rm n}'\zeta\exp(-\zeta/\delta_{\rm n})	& \zeta<0,	   
\end{cases}
\label{eq:Tn.sm}
\end{align}
where $\Gamma_{\rm n}'$ is a parameter that is sufficiently larger than $\Gamma_{\rm n}$.
Although Eq.~\eqref{eq:Tn} is convenient for linear stability analysis, it is inconvenient for numerical simulations; 
if Eq.~\eqref{eq:Tn} was used as the normal traction, areas with $w<0$ would appear.
Since the substrate is solid, negative  values of $w$ are not allowed in a realistic situation.
The modified traction given by Eq.~\eqref{eq:Tn.sm} enables the calculations to avoid such unrealistic solutions.

For the time evolution, we employ a semi-implicit algorithm: 
we use first-order backward and forward finite-difference schemes for the linear and nonlinear parts of Eq.~\eqref{eq:dwdt.1}, respectively.
The $(n+1)$-th step in the calculation of $\tilde{w}$ is given by
\begin{align}
 \tilde{w}^{(n+1)}=\frac{\tilde{w}^{(n)}
-(ik_\beta\tilde{N}_\beta^{(n)}+\tilde{T}_{\rm n}^{(n)})\Delta t}
{1+Dk^4\Delta t},
\end{align}
where $\Delta t$ is the time increment.

\begin{figure}[htb]
 \begin{center}
  \includegraphics[width=7cm,clip]{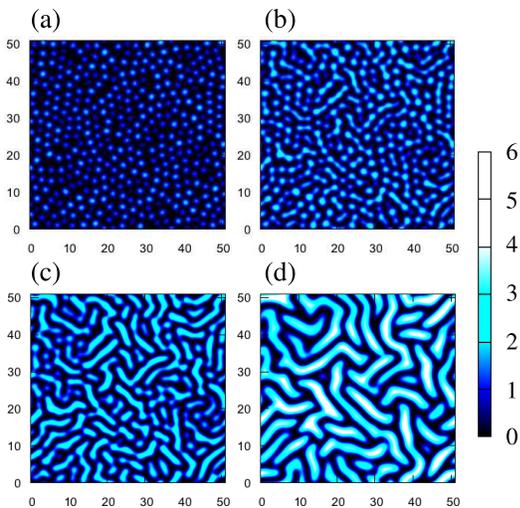}
 \end{center}
 \caption{\label{fig:sm} (Color online)
 Snapshots of numerical simulations at (a) $t=120$, (b) $160$, (c) $200$, and (d) $400$. 
 The color scale illustrates the mid-plane displacement $w$. 
Panels (a)--(b) correspond to (c)--(f) of Fig.~\ref{fig:2-2}, respectively.
The length of a side of a snapshot corresponds to 6.8 mm when the film thickness is $h=0.13$ mm.
}
\end{figure}

Simulated patterns of the displacement $w$ are shown in Fig.~\ref{fig:sm}.
In the initial states, we take $w=0$ plus a small amount of noise, and we impose
periodic boundary conditions on a $256\times 256$ grid system.
The length of a side corresponds to about $6.8$ mm for $h=0.13$ mm.
The parameter values used for Fig.~\ref{fig:sm} are $\varepsilon_0=1.2$, $\nu=0.3$, $\delta_{\rm n}=0.5$,  $\Gamma_{\rm n}=1.4$, and $\Gamma_{\rm n}'=100\;\Gamma_{\rm n}$.

Some characteristics of the snapshots in Fig.~\ref{fig:sm} look similar to those of the experiments in Figs.~\ref{fig:2-2}(c)--(f).
Small bumps appear at an early stage [Fig.~\ref{fig:sm}(a)]. 
The amplitudes of the bumps grow with time, and some bumps coalesce with others [Figs.~\ref{fig:sm}(b) and (c)].
However, the amplitudes continue to grow in the simulations [Fig.~\ref{fig:sm}(d)], which is significantly different from the experiments.
This indicates that our model is not yet adequate to explain the nonlinear effects in the actual experiments.
For more realistic simulations, which will be a focus of future work, stress relaxation and constraints on the total volume of the film should be considered in the model. 

\section{Conclusion}
The objective of this paper has been to understand the emergence of wrinkles at the surface of a coating following the application of an organic solvent. 
The instability at the surface of the layer leads to the emergence of wrinkles. 
We investigated the characteristic lengths of the wrinkles and the characteristic timescale for wrinkle emergence. 
We conclude that (1) buckling due to volume expansion of the layer and (2) delamination of the layer from the substrate are both important for the formation of wrinkles. 
This conclusion is supported by our simple model, which includes the effects of buckling and delamination. 
A linear stability analysis of the model yields a relationship between the wrinkle wavelengths and the timescale for their emergence, 
which agrees reasonably well with our experimental results.

\begin{acknowledgments}
We would like to thank Prof. M. Tokita, Prof. S. Ohta, and Prof. T. Yamaguchi in
Kyushu University, 
R. Ushijima in Ochanomizu University for their fruitful discussions and
suggestions.
We also would like to thank Co. Ishikawa Paint in Osaka for telling us
interesting phenomena observed in coating process in their plant.
This work was supported by JSPS KAKENHI Grant No. 15K04760.
\end{acknowledgments}


\begin{thebibliography}{99}%
	\bibitem{Basu_POC2005} S.~K. Basu, L.~E. Scriven, L.~F. Francis, and A.~V.
	McCormick,
	Prog. Org. Coat. {\bf 53}, 1 (2005).	
	
	\bibitem{Basu_JAPS} S.~K. Basu, L.~E. Scriven, L.~F. Francis, A.~V. McCormick,
	V.~R. Reichert, 
	J. Appl. Polym. Sci. {\bf 98}, 116 (2005).
	
	\bibitem{Thin_film} L. B. Freund and S. Suresh, 
	{\it Thin Film Materials: Stress, Defect and Surface Evolution}
	(Cambridge University Press, Cambridge, 2003).
	
	\bibitem{handbook} E. Takiyama, {\it Polyester Jyushi Handbook}
	(Nikkan Kogyo Sya, Tokyo, 1988) [in Japanese].
	
	\bibitem{Burrell} H. Burrell, 
	Ind. Eng. Chem. {\bf 46}, 2233 (1954).
	
	\bibitem{Gel} T. Tanaka, S.-T. Sun, Y. Hirokawa, S. Katayama, J. Kucera,
	Y. Hirose, and T. Amiya,
	in {\it Molecular Conformation and Dynamics of Macromolecules in Condensed
		Systems}, edited by M. Nagasawa
	(Elsevier, Amsterdam, 1988).
	
	\bibitem{Basu_shiwa} S.~K. Basu, A.~M. Bergstreser, L.~F. Francis, L.~E.
	Scriven, and A.~V. McCormick,
	J. App. Phys. {\bf 98}, 063507 (2005).
	
	\bibitem{tau} R.~Huang and S. H. Im,
	Phys. Rev. E {\bf 74}, 026214 (2006).
	
	\bibitem{Faou} J.~Y. Faou, G. Parry, S. Grachev, and E. Barthel, 
	Phys. Rev. Lett. {\bf 108}, 116102 (2012).
	
	\bibitem{Pan1} K. Pan, Y. Ni, and L. He, 
	Phys. Rev. E {\bf 88}, 062405 (2013).
	
	\bibitem{Ni} Y. Ni, L. He, and Q. Liu, 
	Phys. Rev. E {\bf 84}, 051604 (2011).
	
	\bibitem{Pan2} K. Pan, Y. Ni, L. He, and R. Huang, 
	Int. J. Solids and Struct. {\bf 51}, 3715 (2014).
	
	\bibitem{c-zone1} E. Cerda and L. Mahadevan,  
	Phys. Rev. Lett. {\bf 90}, 074302 (2003).
	
	\bibitem{c-zone2} G.~I. Barenblatt,  
	Adv. Appl. Mech. {\bf 7}, 55 (1962).
	
	\bibitem{c-zone3} K. Park and G.~H. Paulino,  
	Appl. Mech. Rev. {\bf 64}, 060802 (2013).
\end{thebibliography}
\end{document}